\documentclass{article}

\usepackage{arxiv}

\usepackage[square,numbers]{natbib}
\usepackage[utf8]{inputenc} 
\usepackage[T1]{fontenc}    
\usepackage{hyperref}       
\usepackage{url}            
\usepackage{booktabs}       
\usepackage{amsfonts}       
\usepackage{nicefrac}       
\usepackage{microtype}      
\usepackage{lipsum}
\usepackage{graphicx}
\usepackage{listings}
\usepackage{xcolor}
\usepackage{doi}
\usepackage{authblk}

\definecolor{codegreen}{rgb}{0,0.6,0}
\definecolor{codegray}{rgb}{0.5,0.5,0.5}
\definecolor{codepurple}{rgb}{0.58,0,0.82}
\definecolor{codeorange}{rgb}{1,0.4,0}
\definecolor{backcolour}{rgb}{0.95,0.95,0.92}

\lstdefinestyle{mystyle}{
    backgroundcolor=\color{backcolour},   
    commentstyle=\color{codegreen},
    keywordstyle=\color{magenta},
    numberstyle=\tiny\color{codegray},
    stringstyle=\color{codepurple},
    basicstyle=\ttfamily\footnotesize,
    breakatwhitespace=false,         
    breaklines=true,                 
    captionpos=b,                    
    keepspaces=true,                 
    numbers=left,                    
    numbersep=5pt,                  
    showspaces=false,                
    showstringspaces=false,
    showtabs=false,                  
    tabsize=2
}
\graphicspath{ {./images/} }

\newcommand{\graylst}[1]{\colorbox{backcolour}{\lstinline{#1}}}

\newcommand{\PyBryt}[0]{\emph{PyBryt}}

\AtBeginDocument{%
  \providecommand\BibTeX{{%
    \normalfont B\kern-0.5em{\scshape i\kern-0.25em b}\kern-0.8em\TeX}}}

\title{PyBryt: auto-assessment and auto-grading for computational thinking}
\renewcommand{\headeright}{Pyles et al.}
\renewcommand{\undertitle}{}
\renewcommand{\shorttitle}{PyBryt: auto-assessment and auto-grading for computational thinking}
\date{}

\author[1]{Christopher Pyles}
\author[2]{Francois van Schalkwyk}
\author[2]{Gerard J. Gorman}
\author[2]{Marijan Beg} 
\author[1]{Lee Stott}
\author[1]{Nir Levy}
\author[1,3,4,5]{Ran Gilad-Bachrach}
\affil[1]{Microsoft}
\affil[2]{Department of Earth Science and Engineering,  Imperial College London, United Kingdom}
\affil[3]{Biomedical Engineering Department, Tel-Aviv University}
\affil[4]{Edmond J. Safra Center for Bioinformatics, Tel-Aviv University}
\affil[5]{Sagol School of Neuroscience, Tel-Aviv University}
\begin{document}

\maketitle

\begin{abstract}
We continuously interact with computerized systems to achieve goals and perform tasks in our personal and professional lives. Therefore, the ability to program such systems is a skill needed by everyone. Consequently, computational thinking skills are essential for everyone, which creates a challenge for the educational system to teach these skills at scale and allow students to practice these skills. 
To address this challenge, we present a novel approach to providing formative feedback to students on programming assignments. Our approach uses dynamic evaluation to trace intermediate results generated by student's code and compares them to the reference implementation provided by their teachers. We have implemented this method as a Python library and demonstrate its use to give students relevant feedback on their work while allowing teachers to challenge their students' computational thinking skills.

\end{abstract}


\section{Introduction}

In her seminal paper, Wing~\cite{wing2006computational} argued that computational thinking is a skill that is not limited only to computer programming but that it has a much broader impact. Furthermore, Wing argued that computational thinking should be taught to any college freshman and should not be limited to a computer science major. Some policymakers argue that computational thinking should be taught in the compulsory education system. In Finland, for example, ``algorithmic thinking'' is a mandatory topic for first-grade students~\cite{bocconi2016developing}. Accordingly, computational thinking should be taught at a massive scale~\cite{CC2020}, with higher education playing a critical role in ensuring graduates have the necessary industry-transferable digital skills for the workplace~\cite{ali2020online}. This makes tools for automated guiding of students as well as for providing automated formative (auto-assessment) and summative (auto-grading) feedback of paramount importance~\cite{taras2005assessment, winstone2020need}. Indeed, the need for such tools has been identified over 60 years ago~\cite{hollingsworth1960automatic}. The Adoption of Auto-Grading (AG) and Auto-Assessment (AA) tools has seen significant growth in recent years, and it is now widely recognized as an enabler of global-scale education pedagogy and academic support system~\cite{haldeman2018providing,eicher2021components, nabil2021evalseer}. The recent growth of remote learning challenges educators to provide more support for unattended and asynchronous learning.

Many AA \& AG tools are built around a set of unit tests~\cite{forsythe1965automatic, edwards2008web, sarwate2018grading, blank2019nbgrader, manzoor2020auto}. This method for dynamic evaluation provides a pass or fail score for the entire student's code or for subsets of the code for which the API was specified in advance. This means that, to enable fine-grained feedback, the instructor has to provide a detailed description of the API -- a process that has been referred to as unit test ``scaffolding'', or microlearning~\cite{Skalka2021}. However, in many cases, this process deprives the students of a better discovery-oriented learning experience that a tutor and not an automatic unit test can provide.

As an example, a student who is asked to write a program to compute the median of elements in a list may get a scaffolded code skeleton, similar to the one presented in Listing~\ref{lst:scaffolded-test}, and asked to complete missing parts.

\begin{lstlisting}[language=Python, caption=Scaffolded assignment example,  label={lst:scaffolded-test}]
S = [8.0, 5.0, -1.0, 3.0, 5.0, 2.25]
sorted_S = ...
size_of_S = ...
middle = ...
is_size_of_S_even = ...
if is_size_of_S_even:
    median = ...
else:
    median = ...

print("The median is ", median)
\end{lstlisting}

Due to the limitation of the automated evaluation tools, the instructor provides the student with the architecture of the solution. Thus, the student is practising basic programming skills and is not challenged with fundamental computational thinking tasks such as decomposing the problem and considering multiple possible solutions. This may be a good exercise for novice students doing their first coding steps, but it is ineffective when practising student's ability to come up with the architecture by themselves. On the other hand, a much less restrictive exercise for computing the median may ask the student to complete the code as presented in Listing~\ref{lst:open-test}, allowing multiple implementation patterns and approaches. However, this format makes it harder to generate constructive feedback automatically.
\begin{lstlisting}[language=Python, caption=Open assignment example, label={lst:open-test}]
def median(S):
    ...
\end{lstlisting}

Static code analysis has been proposed as another solution for AA \& AG~\cite{yulianto2014automatic, striewe2014review}. Unlike unit-test-based analysis, these methods are not limited by the interface of the solution. Instead, this is done by analyzing the student's source code or the byte-code and comparing it against a set of rules specified by the instructor. However, defining the rules to be checked is challenging, and therefore most instructors would find it hard to use. Recently, deep-learning-based approaches were presented for comparing students' solutions to each other and to reference implementations~\cite{bhatia2016automated, de2019convolutional}. However, \citet{bhatia2016automated} discuss only the correction of syntax errors while \citet{de2019convolutional} focus only on auto-grading.

In this work, we present a new method to assess student's work which we have implemented in the \PyBryt{} library which is available at \url{https://github.com/microsoft/pybryt}.  In \PyBryt{}, the instructor is providing one or several \emph{reference implementations} of the given assignment. The teacher annotates these reference implementations by marking both expected and not expected intermediate values to be computed by the student. The instructor can also specify feedback to be given to the student depending on whether these annotated values are present or absent. The \PyBryt{} library executes the student's code and traces intermediate values to the ones marked by the instructor. Based on the matching between the observed values and expected values, feedback is given to the student. This approach allows controlling the freedom given to students: the instructor can provide fine-grained steps and trace the student's ability to follow instructions, or alternatively, the student can work on an open problem by breaking it into sub-tasks and practising computational thinking.

In Section~\ref{sec:reference-implementations} we present the concept of reference implementations in \PyBryt{} and how they are used to asses students code. In Section~\ref{sec:examples} we present several examples for the use of \PyBryt{} for providing feedback to students and for plagiarism detection. We conclude with a discussion in Section~\ref{sec:summary}. 

\section{Reference Implementations}
\label{sec:reference-implementations}
To provide feedback to students or to grade their solutions, the instructor specifies their expectations. In \PyBryt{}, expectations are defined using a \emph{reference implementation} (RI) -- a set of conditions expected from the students' code. RI's are constructed by annotating series of values expected to be present in the student's solution, even if not explicitly specified in the guidelines for the assignment. For example, suppose the student is asked to implement the median function. Apart from the correct return value, the instructor may expect the student to sort the list of numbers in their implementation and calculate the list's length to determine if it has an odd or even number of items. Therefore, the instructor annotates these values in RI. In \PyBryt{}, this is done as shown in Listing~\ref{lst:median}. When the student's code is executed in the \PyBryt{} environment, all intermediate values that are created by the student's solution are scanned to find the values marked by the instructor in RI. The instructor has the option to provide success-messages to be presented to the student if their code computes the expected values and failure-messages to present if the values are missing. Therefore, the instructor can provide formative feedback that can encourage and focus the students on ways to improve.

\begin{lstlisting}[language=Python, caption=Median example annotated by the instructor, label={lst:median}]
import pybryt

def median(S):
    sorted_S = sorted(S) 
    pybryt.Value(sorted_S, 
                  success_message = "Great: The list is sorted correctly.", 
                  failure_message = "Advice: The list was not sorted.")
    
    size_of_S = len(S) 
    pybryt.Value(size_of_S, 
                  success_message = "Great: The size of the list is computed.", 
                  failure_message = "Advice: The size of the list is not computed to determine if the number of elements is odd or even.")
    
    middle = (int)(size_of_set / 2) 
    is_size_of_S_even = (size_of_set % 2) == 0

    if is_size_of_S_even:
        return (sorted_S[middle-1] + sorted_S[middle]) / 2
    else:
        return sorted_S[middle]
\end{lstlisting}

It is important to note that only the instructor adds annotations to their RI. Students are not required to modify their code. Instead, \PyBryt{} executes students code in the debugger framework and traces the memory footprint to search for the values marked in the RI. All intermediate values are extracted regardless of variable names or function names. Accordingly, the teacher can design their assignments without "scaffolding" or using microlearning~\cite{Skalka2021}. Furthermore, the computational complexity of the solution can be evaluated using the number of steps taken to run the program, which is a consistent measure that is independent of the hardware on which the program is executed or other programs running on the computer simultaneously. \PyBryt{} allows instructors to specify in their RI the expected computational complexity of the solution and provide students feedback when they do or do not match these expectations.


To demonstrate the way \PyBryt{} works consider the student code in Listing~\ref{lst:median-student-1} as a solution to the median assignment. The solution does compute the median correctly; however, it will trigger two failure messages since expected intermediate values are not computed. On the other hand, the solution in Listing~\ref{lst:median-student-2} will trigger two success messages even though the student did not use the same variable names and sorted the list in-place as opposed to the RI where new memory is allocated. Nevertheless, \PyBryt{} recognizes that the required values were computed by the student and triggers success messages.

\begin{lstlisting}[language=Python, caption={Median example - student implementation with numpy. Although this solution computes the median correctly, \PyBryt{} can identify that the student uses a library call to compute the median.}, label = {lst:median-student-1}]
def median(S):
    import numpy as np
    return np.median(S)
\end{lstlisting}

\begin{lstlisting}[language=Python, caption=Median example - student implementation with in memory sorting, label = {lst:median-student-2}]
def median(T):
    T.sort()
    middle = (int)(len(T) / 2) 
    is_set_size_even = (len(T) % 2) == 0

    if is_set_size_even:
        return (T[middle-1] + T[middle]) / 2
    else:
        return T[middle]
\end{lstlisting}

\PyBryt{} traces values computed by students solutions even when they are not assigned to variables. Note that the second value the instructor marked in the median assignment is the length of the list, however, in Listing~\ref{lst:median-student-2}, this value is computed but never assigned to a variable. Nevertheless, \PyBryt{} detects that the value was computed and triggers the success message. To achieve that, \PyBryt{} first analyzes the program by parsing its Abstract Syntax Tree (AST) and assigns each intermediate result to a variable. Hence, \PyBryt{} first does static evaluation of the student's solution and creates a version of the code where every intermediate result is assigned to a variable. Next, it runs the code in debug mode to trace every value assigned to a variable while executing the code.

\subsection{Logic and temporal rules over values of interest}
In some cases, the instructor may be interested in defining logical rules over the annotated values. For example, consider an assignment where the student is expected to find the maximum value in a list. The instructor may wish to verify that the student does not sort the list since this increases the complexity of the solution. Therefore, the instructor may define the RI as shown in Listing~\ref{lst:max-value}.
\begin{lstlisting}[language=Python, caption=Finding the maximal value in a list, float=bt,  label = {lst:max-value}]
import pybryt

def max_of_list(S):
    sorted_S = sorted(S) 
    pb_sort = ~pybryt.Value(sorted_S)
    pb_sort.failure_message = "Advice: sorting the list adds unnecessary complexity"
    size_of_set = len(S) 
    return(sorted_S[len(S)-1])
\end{lstlisting}

In the RI presented in Listing~\ref{lst:max-value}, the logical \emph{not} rule is introduced using the \graylst{\~} symbol. This means that \graylst{pb\_sort} is satisfied when the sorted version of \graylst{S} is not encountered, and the failure message is triggered only when the value is present. \PyBryt{} supports several logical rules: the \emph{or} operator \graylst{|}, the \emph{and} operator \graylst{\&}, and the not operator \graylst{\~}. Logical rules are useful in many settings. For example, if the student can use several data structures to store the data, an or-statement between the different values can be used. In some cases, it may be easier for the instructor to support such cases using multiple RI's as discussed in Section~\ref{subsec:multy}.

Besides logic operations, \PyBryt{} also supports two temporal operators: \graylst{before} and \graylst{after}. If we assume values \graylst{v1} and \graylst{v2}, \graylst{v_before = v1.before(v2)} will trigger the success message if, in the student's solution, the first time \graylst{v1} is satisfied precedes the first time \graylst{v2} is satisfied. Similarly, \graylst{v_after = v1.after(v2)} expects \graylst{v1} to occur after \graylst{v2}. Since this is a common use-case, \PyBryt{} also supports \emph{collections}. A collection is a list of annotated values expected to be present. The instructor can specify whether the student's solution should generate these values in the same order as in the RI or the order in which the values are generated is not a requirement. Moreover, the result of a logical or temporal operation is another Value object, allowing rules to be chained to create more complex conditions when needed.

\subsection{Multiple reference implementation}\label{subsec:multy}
When evaluating students code, \PyBryt{} first runs student's implementation, traces and stores every value that is being computed, and finally compares them to those marked in the RI. Since the values created by the student's implementation are stored, they can be compared to several RI's. A single RI is restrictive since there may be multiple algorithms for solving a given problem. For the median example, one possible solution is to sort the list and find the item in the middle of the sorted list (or the average of the two middle items if the number of items is even). This solution has time complexity of $O\left( n \log \left( n\right) \right)$. However, there are more efficient solutions, such as the randomized quick-select algorithm~\cite{hoare1961algorithm} or the deterministic pick-pivot algorithm~\cite{blum1973time}. These solutions are very different from the sort-based solution and therefore generate a different memory footprint. Therefore, \PyBryt{} allows different possible solutions to be covered by separate RI's.

Since extracting the memory footprint from students' solutions may be time-consuming, \PyBryt{} generates an object that stores all values so that comparison to multiple RI's is more efficient. In addition, the RI can be "compiled" into an object that stores all the required values and logical rules to further reduce the time.

Another setting in which the use of multiple RI's is useful is when a mistake in an early stage may lead to a deviation in the computation. For example, consider the assignment that asks the student to count the number of times a substring matching the pattern 'ab?d' appears in the text such that the matching is case insensitive. A student that did not convert the text to lower-case will not match many of the intermediate results expected and therefore may get many error messages. However, if this happens to be a common mistake, the teacher may create a RI for this type of ``solution''. It will capture the error and be able to test other aspects of the solution. 

Having multiple RI's allows additional types of assignments that can be given to students to practice their computational thinking skills. For example, an assignment can ask students to implement two types of sorting algorithms. Note that \PyBryt{} can verify that the solutions are different by verifying that they match different RI's since they generate different memory footprints.

\subsection{Computational complexity}
Since \PyBryt{} runs students' solutions in debug mode, it also counts the number of steps taken during the execution of the code. This can be useful when students are learning about computational complexity. Students can try different implementations and see how their choices affect the number of steps taken. This can be used in multiple ways. The teacher can specify the required complexity or ask students to implement both bubble-sort and merge-sort and draw a chart showing the number of steps it took for their solution to complete as a function of the input size. 

Counting steps can be used even for learning fine-grain optimization techniques. For example, students can be asked to find the prime numbers in a given list of integers. In this assignment, caching can be used to accelerate the computation. Therefore, students can be asked to find an efficient solution, and \PyBryt{} can be used to provide them with feedback about the efficiency. This can also be used for gamification by creating a class challenge to find a solution that uses the fewest instructions. Note that since students may be using different computers, measuring the time to execute may give an unfair advantage to students that have access to faster computers, but  \PyBryt{} can measure the complexity independent of hardware.

\subsection{Programming style}
\PyBryt{} can be used to provide feedback helping students to improve their programming style. Consider, for example, an assignment where the student is asked to implement a function that takes a list of numbers and returns a list of these numbers squared. In Listing~\ref{lst:square}, two possible solutions are presented. Both \graylst{solution1} and \graylst{solution2} in Listing~\ref{lst:square} are correct. However, \graylst{solution1} using list-comprehension is more readable and thus preferred. To see which solution the student used, it is possible to check the intermediate values. When using a loop, as \graylst{solution2} does, the \graylst{result} list is gradually built and therefore it starts as an empty set, followed by having the value \graylst{[1]}, then \graylst{[1, 4]} and finally \graylst{[1, 4, 9]}. However, list comprehension creates the list in a single step. Therefore, the value \graylst{[1, 4]}, as an example, will never be observed, although the value \graylst{[1, 4, 9]} will, and the instructor can mark the value \graylst{[1, 4]} to identify the use of a for-loop. It should be noted that static methods for evaluating programs can also be used to identify different solutions. However, the benefit of using \PyBryt{} for this task is the fact that the same tool and syntax can be used in many scenarios, which makes it easier for the instructor to use.

\begin{lstlisting}[language=Python, caption=Two methods for computing the square of every item in the list, label = {lst:square}]
def solution1(lst):
    result = [x*x for x in lst]
    return result

def solution2(lst):
    result = []
    for x in lst:
        result.append(x*x)
    return results
\end{lstlisting}

\subsection{Tolerances and Invariances}
A possible drawback of using \PyBryt{}'s approach is tolerance to slight fluctuations in results. For example, when using real numbers, small numerical errors may occur due to the finite precision of the computation, e.g. the statement \graylst{1 == ((0.3 * 3) + 0.1)} evaluates to False in Python 3.8.5. To overcome this issue, \PyBryt{} supports tolerances when comparing numerical values. Both relative and absolute tolerances are supported in a similar fashion to NumPy's \graylst{isclose} function ~\cite{oliphant2006guide}.

Similar problems may be encountered when working with strings where the instructor may be invariant to using lower-case vs upper-case letters. \PyBryt{} supports invariants by specifying them when declaring a value. Therefore, if the RI declares:\\
\graylst{pybryt.Value("hello", invariants=[string_capitalization])} \\
then it will match the strings "Hello", "HELLO", and "HeLLo".  Other invariants supported include list-permutation and matrix-transpose. Furthermore, the instructor can implement additional invariances.

\subsection{Limitations}
It is important to note that \PyBryt{} may not be suitable for any possible assignment. For example, the current version of \PyBryt{} is not designed to test solutions that use parallelism or solutions that operate on large memory objects. It also induces running time overheads, and therefore solutions that are slow to run might become excessively slow when run through \PyBryt{}. Random algorithms may also be hard to evaluate with \PyBryt{} although this problem can be avoided by fixing the seed of the random number generator.

\section{Examples}
\label{sec:examples}
In this section, we present several examples of the use of \PyBryt{}.

\subsection{Counting words}
Here, we consider an assignment in which students are asked to write a function that counts the number of words in a string. The instructor can provide students with the signature of the required function (Listing~\ref{lst:Problem}) and ask them to complete it.

\begin{lstlisting}[language=Python, caption=Count words assignment without scaffolding., label={lst:Problem}]
def count_words(text: str) -> int:
    ...
\end{lstlisting}

A student may implement the \graylst{count\_words} function by splitting the string using built-in methods and then count the number of items in the resulting list as demonstrated in Listing~\ref{lst:Function}.

\begin{lstlisting}[language=Python, caption=Count words implementation, float=tb, label={lst:Function}]
def count_words(text: str) -> int:
    fragments = text.split()
    length = len(fragments)
    return length
\end{lstlisting}   

To create a RI, the instructor may want to trace: (i)~the list of words created by splitting the text (\graylst{fragments}), and (ii)~the count of words as a Python int (\graylst{length}). This RI is shown in Listing~\ref{lst:counting-ref}. The key thing to observe is the ease with which the instructor can create a RI. To check the student's solution, it can be executed with different values as inputs as demonstrated in Listing~\ref{lst:Check Implementation}.

\begin{lstlisting}[language=Python, caption=Count words reference implementation, float=tb, label={lst:counting-ref}]
def count_words(text: str) -> int:
    # Split the text by spaces
    fragments = text.split()
    pybryt.Value(fragments, failure_message="It may help to first split the string into a list of words", success_message="Great, the string was converted into a list of words")

    # Count the number of items in the result
    length = len(fragments)
    pybryt.Value(length,
                  success_message="Great, identified the number of words in the string",
                  failure_message="Cannot find a count of the number of words")
    return length
\end{lstlisting}

\begin{lstlisting}[language=Python, caption=Unit tests for count words assignment, float=tb, label={lst:Check Implementation}]
assert count_words("Waltz, bad nymph, for quick jigs vex.") == 7
assert count_words("The five boxing wizards-- \njump quickly!") == 6
assert count_words("Sphinx of black quartz,          judge my vow.") == 7
\end{lstlisting} 

\subsection{Insertion Sort}
In this example, we consider an assignment in which students are asked to implement insertion sort. With \PyBryt{}, the instructor does not need to create a scaffolded version of the solution, instead, the student can be asked to implement a function: \graylst{insertion\_sort} that takes a list of integers as its input and returns the list sorted in ascending order using insertion sort.

\begin{lstlisting}[language=Python, caption={A correct implementation of an insertion sort assignment}, float=tb, label={lst:Correct Answer}]
def insertion_sort(x: list[int]) -> list[int]:
    for i in range(1, len(x)):
        current = x[i]
        j = i - 1
        while j >=0 and current < x[j] :
            x[j+1] = x[j]
            j -= 1
        x[j+1] = current
    return x
\end{lstlisting}     

\begin{lstlisting}[language=Python, caption={An incorrect implementation of an insertion sort assignment, note that the outer loop iterates to `len(x)-1` instead of `len(x)`. }, float=tb, label={lst:Wrong Answer}]
def insertion_sort(x: list[int]) -> list[int]:
    for i in range(1, len(x)-1):  # MISTAKE: len(x)-1 instead of len(x)
        current = x[i]
        j = i - 1
        while j >=0 and current < x[j] :
            x[j+1] = x[j]
            j -= 1
        x[j+1] = current
    return x
\end{lstlisting}   

Listing~\ref{lst:Correct Answer} presents a possible correct answer to this assignment while Listing~\ref{lst:Wrong Answer} presents an incorrect solution in which the student's outer loop does not iterate to the end (\graylst{len(x)-1} instead of \graylst{len(x)}). The instructor would like to distinguish the correct solution from the incorrect one and provide formative feedback. A unit-test can be used to tell the correct implementation from the erroneous one as demonstrated in Listing~\ref{lst:Unit Test Feedback}. From this feedback, both the student and educator can only see if the final solution is correct. However, even if the solution is correct, it cannot determine whether insertion sort was used or not. Moreover, if the student did not implement the code correctly, the feedback does not provide information on how to improve the code. Therefore, this method provides little value both in the formative feedback it provides and when used to provide summative feedback.

\begin{lstlisting}[language=Python, caption={Unit test feedback for the insertion sort assignment. Note that the feedback is non-constructive from student's perspective} , float=tb, label={lst:Unit Test Feedback}]
x = [55, 111, -33, 65, 1001, -362, 451]
assert insertion_sort(x) == list(sorted(x))
-----------------------------
AssertionError  Traceback (most recent call last)
<ipython-input-6-ad50795e9ab5> in <module>
      1 x = [55, 111, -33, 65, 1001, -362, 451]
----> 2 assert insertion_sort(x) == list(sorted(x))

AssertionError: 
\end{lstlisting}   

In contrast to non-informative feedback provided by unit-tests, using \PyBryt{}, the solution in Listing~\ref{lst:Correct Answer} can be converted into a RI such that the feedback for the incorrect assignment Listing~\ref{lst:Wrong Answer}  would be more informative as demonstrated in Listing~\ref{lst:PyBryt Feedback}.

\begin{lstlisting}[language=Python, caption={\PyBryt{} feedback for the incorrect implementation of the insertion sort assignment}, float=tb,label={lst:PyBryt Feedback}]
with pybryt.check(PyBryt_reference(1, 1)):
    insertion_sort([55, 111, -33, 65, 1001, -362, 451])

REFERENCE: exercise-1_1
SATISFIED: False
MESSAGES:
  - ERROR: Hmmm... It looks like the algorithm does not perform insertion sort.
  - ERROR: Please make sure the outer loop iterates from the second element to the last.
  - SUCCESS: Great! The inner loop iterates towards left.
  - ERROR: The function returns a wrong solution.
 \end{lstlisting} 
 
From \PyBryt{}'s feedback, the student can see that although the inner loop is correct, the outer loop is wrong. Because of this, the algorithm does not perform sort as expected. This allows the student to see if their final solution is correct, and \PyBryt{} points the student to the ``outer loop'' to correct the mistake. Similarly, the instructor does not have to go into the source code and find an error. Instead, the educator can resort to \PyBryt{}'s feedback. As such, \PyBryt{} gives much more detailed feedback emphasizing all the points student's function does right and where it can be improved. This way, we can ensure the function performs insertion sort and not some other sorting algorithm. To demonstrate that consider a student that instead of implementing insertion-sort calls Python's \graylst{sorted} function to sort the list as shown in Listing~\ref{lst:Problem}. This solution will pass the unit test since the list is sorted correctly. However, \PyBryt{} can recognize that the ingredients of insertion-sort are missing while the result is correct as demonstrated in the feedback provided by \PyBryt{} in  Listing~\ref{lst:Solution}.

\begin{lstlisting}[language=Python, caption={\PyBryt{} feedback for the insertion-sort assignment when a student implement a sorting algorithm which is not insertion-sort}, float=tb, label={lst:Solution}]
with pybryt.check(PyBryt_reference(1, 1)):
    insertion_sort([55, 111, -33, 65, 1001, -362, 451])

REFERENCE: exercise-1_1
SATISFIED: False
MESSAGES:
  - ERROR: Hmmm... It looks like the algorithm does not perform insertion sort.
  - ERROR: Please make sure the outer loop iterates from the second element to the last.
  - ERROR: Please make sure the inner loop iterates towards left.
  - SUCCESS: Amazing! The function returns the correct solution.
\end{lstlisting} 

Therefore, \PyBryt{} has the potential of saving time for instructors when marking the submissions of entire classes. On the other hand, the student can see all the steps that they did right, which helps with the learning experience and it is more encouraging thus enhancing the overall student experience and increasing student satisfaction. When implemented as a self-assessment tool, students can receive immediate encouraging and constructive feedback which may create benefits from gamification perspectives.

\subsection{Plagiarism Detection}

Some accounts suspect that at least ~20\% of students are involved in programming plagiarism, even at the top universities.\footnote{\url{https://www.nytimes.com/2017/05/29/us/computer-science-cheating.html}} Others suggest that the numbers might be even higher: in a survey of 138 students, 72.5\% admitted to have plagiarized at least once in a programming assignment~\cite{sraka2009source}.
As a result, many methods have been developed for detecting plagiarism.
To illustrate that, we note that at the time of writing this paper,\footnote{in August 2021} Google Scholar\footnote{\url{https://scholar.google.com/}} reports that 21000 papers are matching the query `plagiarism programming assignments', 7500 of them were published since 2017, and 2650 since 2020. Scholars also reported that reducing plagiarism does improve students competency. For instance, \citet{pawelczak2018benefits} reported that since the introduction of automatic plagiarism detection, students scores improved by 8.6\%. 

In an attempt to conceal plagiarism, students use many techniques ranging from changing variable names and adding blank lines to breaking functions and adding non-functional code~\cite{petrik2017source}. \PyBryt{} can act as a plagiarism detection tool by automatically converting students' submissions into RI's and comparing other students submissions to it. Since \PyBryt{} compares only the values of variables, it is resilient to obfuscation techniques such as variable renaming, the addition of redundant code, changes to the order of computation that does not result in functional changes and more.

To use PyBryt as a plagiarism detection mechanism we note that if a solution was plagiarized then both the original version of the code and the copied version of the code will share most intermediate values. This is true even if variable names where changed or function names were changed. Moreover, even if a sophisticated student changed the order of calls to functions in places where it would not have consequences on the results then PyBryt can detect the plagiarism. One potential problem is that some intermediate values are very common and are not a reliable signal for plagiarism. To circumvent this problem, we propose using a method akin to TF-IDF to score the 'suspiciousness' of finding a value matching in the code submitted by separate students.If a value is common in the solutions of many students, then this value is not a significant marker for plagiarism.

\section{Summary and Conclusions}
\label{sec:summary}

\citet{CC2020} argued that learning should include both knowledge, skill, and disposition. More precisely, \citet{Corbett2001} found that in learning programming, students benefit from explicit and in-context guidance for coding assignments, compared to other means of feedback. This is especially important in light of the big gap in the faculty-to-teacher ratio in different colleges: \citet{Ertek2018} presented in their study a comprehensive benchmark of the top 100 U.S. universities and found that the faculty-to-student ratio is a significant competitiveness factor.

In this work, we presented \PyBryt{}, which uses a new approach to providing feedback to students on their coding assignments. \PyBryt{} compares students solutions to RI's by checking for the presence or absence of critical values, as marked by the instructors. Thus \PyBryt{} is easy to use yet does not require the teacher to provide scaffolded exercises. 

A side benefit of our approach is the ability to detect plagiarism. Plagiarism was identified as a concern in programming education since the 70's~\cite{traxler1970plagiarism, ottenstein1976algorithmic}, and continues to worry educators today~\cite{Novak2019}. As \PyBryt{} follows memory evolution during run time, it can detect similarity between programs - even if the source code is different, e.g. different variable names or redundant code.

This paper presents \PyBryt{} and the way it is implemented. We are currently working on evaluating \PyBryt{} in different pedagogical settings to measure its contribution to the learning process of students and the effort required from educators to use it. \PyBryt{} can be extended in many ways. In its current form \PyBryt{} supports only programs in python while it can be extended to support many other programming languages. Furthermore, it can support cross-lingual assessments since the intermediate results computed are invariant to the programming language used in many cases.

Moreover, \PyBryt{} may be used beyond the classroom for tasks such as code-refactoring. Code-refactoring is a challenging task even for seasoned engineers. \PyBryt{} can support this process by converting the original code into a reference implementation and testing the refactored code against it. This allows the detection of errors in the new implementation even when the code's internal and/or external structure does not match the original one.

\bibliographystyle{ACM-Reference-Format}
\bibliography{references}

\end{document}